\documentclass[fp,twocolumn]{jpsj3}
\usepackage{txfonts}
\usepackage{url}
\usepackage{xcolor}

\usepackage{bm}
\usepackage{amsfonts}
\usepackage{physics}
\usepackage{float}
\usepackage{here}
\usepackage{comment}
\usepackage{ulem}

\usepackage{color}

\title{Quantitative Analysis of the Effectiveness of Mid-anneal Measurement in Quantum Annealing}

\author{Keita Takahashi$^{1}$ and Shu Tanaka$^{1,2,3,4}$\thanks{shu.tanaka@keio.jp}}
\inst{$^1$Graduate School of Science and Technology, Keio University, Yokohama, Kanagawa
223-8522, Japan \\
$^2$Department of Applied Physics and Physico-Informatics, Keio University, Yokohama, Kanagawa 223-8522, Japan \\
$^3$Keio University Sustainable Quantum Artificial Intelligence Center (KSQAIC), Keio University, Minato, Tokyo 108-8345,
Japan \\
$^4$Human Biology-Microbiome-Quantum Research Center (WPI-Bio2Q), Keio University, Shinjuku, Tokyo 108-8345, Japan}

\abst{
Quantum annealing is a promising metaheuristic for solving constrained combinatorial optimization problems. However, coefficient tuning difficulties and hardware noise often prevent optimal solutions from being properly encoded as the ground states of the problem Hamiltonian. This study investigates mid-anneal measurement as a mitigation approach for such situations, analyzing its effectiveness and underlying physical mechanisms. We introduce a quantitative metric to evaluate the effectiveness of mid-anneal measurement and apply it to the graph bipartitioning problem and the quadratic knapsack problem. Our findings reveal that mid-anneal measurement is most effective when the energy difference between desired solutions and ground states is small, with effectiveness strongly governed by the energy structure. Moreover, analysis of fully-connected Ising models demonstrates that the effectiveness of mid-anneal measurement persists with increasing system size, indicating its scalability and practical applicability to large-scale quantum annealing.}

\begin{document}
\maketitle

\section{Introduction}
Constrained combinatorial optimization problems arise in various fields including finance~\cite{Grant2021benchmarking, Hidaka2023correlation}, materials science~\cite{Hernandez2017enhancing, Micheletti2021polymer, utimula2021quantum, sampei2023quantum, hatakeyama2023extracting, tucs2023quantum, xu2025quantum}, and transportation~\cite{Inoue2021traffic, Bao2021-a, Bao2021-b, haba2022travel, bao2023ising, kanai2024annealing}. These problems are generally characterized by an exponential increase in the number of candidate solutions with increasing problem size. Quantum annealing has attracted attention as a promising method for solving such problems~\cite{Kadowaki1998quantum, das2008colloquium, chakrabarti2023quantum}. To find optimal solutions using quantum annealing, constrained combinatorial optimization problems must be formulated as ground-state search problems of Ising models. The Ising model, in which the energy is determined by interactions between $\pm 1$-valued spin variables and local magnetic fields acting on spin variables, provides a general framework for representing many combinatorial optimization problems~\cite{Lucas2014ising, Tanaka2017quantum, tanahashi2019application}.
Therefore, the problem Hamiltonian $H_\mathrm{c}$ is typically formulated using a penalty function. Consider an optimization problem with an objective function $f(\bm{s})$ to be minimized, subject to equality constraints $c_k(\bm{s}) = 0$ for $k = 1, 2, \ldots, K$, where $\bm{s}$ is a vector of spin variables and $K$ is the number of constraints. By incorporating the constraints as penalty terms, the problem Hamiltonian $H_\mathrm{c}$ can be constructed as follows:
\begin{equation}
    H_\mathrm{c} 
    = f(\bm{s})
    + \frac{\mu}{2} \sum_{k=1}^K \left[ c_k(\bm{s}) \right]^2,
\label{eq: penalty func}
\end{equation}
where $\mu$ is the coefficient of the constraint term $\sum_{k=1}^K \left[ c_k(\bm{s}) \right]^2$. Note that $\mu$ is a penalty coefficient that takes a positive value. For solutions that satisfy all constraints, the constraint terms $\left[ c_k(\bm{s}) \right]^2$ attain their minimum value (zero) for all $k$, while for solutions that violate any constraints, a penalty is imposed. Additionally, recent studies have explored augmented Lagrangian functions that incorporate linear terms weighted by coefficients $\lambda_k$, in addition to the penalty function as shown below~\cite{Cellini2024qal, Djidjev2023quantum, Djidjev2023logical}:
\begin{equation}
    H_\mathrm{c} 
    = f(\bm{s})
    + \frac{\mu}{2} \sum_{k=1}^K \left[ c_k(\bm{s}) \right]^2
    - \sum_{k=1}^K \lambda_k \left[ c_k(\bm{s}) \right].
\label{eq: AL func}
\end{equation}
The inclusion of linear terms allows for better control over constraint satisfaction and can improve convergence behavior by incorporating directional information of the constraint violations~\cite{Cellini2024qal}.

In this study, we focus on the augmented Lagrangian approach for constrained combinatorial optimization problems due to its demonstrated superior performance over standard penalty methods in quantum annealing~\cite{Cellini2024qal, Djidjev2023quantum, Djidjev2023logical}. When the penalty coefficients become excessively large, the required precision may exceed the hardware capabilities, preventing the correct encoding of the intended Hamiltonian~\cite{hattori2025impact}. Therefore, the augmented Lagrangian method is particularly effective in practice. The augmented Lagrangian method achieves better solution quality while requiring smaller penalty coefficients than traditional approaches, thereby mitigating numerical precision issues inherent in quantum hardware~\cite{Djidjev2023quantum, Djidjev2023logical}.
To encode optimal solutions as the ground states of the problem Hamiltonian and obtain them with high probability, tuning the constraint coefficients $\mu$ and $\lambda_k$ to appropriate values is essential~\cite{Ohzeki2020breaking, Villar2022analyzing, Djidjev2023quantum,  Quinton2025quantum}. Specifically, coefficient tuning is the process of adjusting the coefficients of constraint terms to balance constraint satisfaction with objective function minimization. This process involves a strict trade-off: if the coefficients are too small, the impact of constraint terms is insufficient, leading to a ground state that violates constraints. Conversely, if they are excessively large, the constraint terms dominate the energy landscape, diminishing the relative contribution of the objective function and making the search for the optimal solution inefficient. This difficulty arises from the need to search over a multi-dimensional hyperparameter space defined by the coefficients $\mu$ and $\lambda_k$, whose size increases with the number of constraints. In general, when multiple parameters are involved in constraint handling, estimating their suitable values becomes extremely difficult~\cite{Yin2024penalty}. Consequently, tuning to inappropriate values often leads to a scenario where the true optimal solution is encoded not as the ground state of the Hamiltonian.

Furthermore, in practical quantum annealing applications, the presence of analog hardware noise inherent to quantum annealers poses a serious issue for encoding optimal solutions. Quantum annealers consist of physical qubits interconnected by tunable couplers that implement the required interaction topology for a given optimization problem. Due to the analog nature of these hardware components, control imperfections are inevitable in practical implementation. This noise manifests as uncertainties in control parameters such as interactions and local magnetic fields, originating from various sources ranging from external environments to unused couplers\cite{Chancellor2022error, Willsch2022benchmarking}.
When the problem Hamiltonian is subject to such noise, even small perturbations may significantly alter the energy landscape and the stability of optimal solutions. In systems with complex energy landscapes, the ground state configuration can become highly sensitive to these parameter fluctuations. One prominent manifestation of this issue is $J$-chaos, particularly in difficult combinatorial optimization problems such as finding the ground state of spin glasses\cite{Nifle1992new, Nifle1998chaos, Katzgraber2007temperature, Albash2019analog, Pearson2019analog}. $J$-chaos is a phenomenon where the ground state of the system becomes unstable even under weak noise. $J$-chaos, named after the sensitivity to interaction parameters, arises when small changes in these coupling strengths cause dramatic rearrangements of the energy spectrum, leading to different ground state configurations. As a result, the ground state may shift to an unintended configuration. Regarding this effect, it has been reported that under a fixed noise level, the probability that optimal solutions are properly encoded as ground states of the Hamiltonian decreases exponentially with increasing problem size\cite{Albash2019analog}. This scaling law suggests that $J$-chaos is a serious phenomenon for solving large-scale problems. Furthermore, experimental verification using D-Wave Systems' quantum annealers has also demonstrated the severity of $J$-chaos in large-scale problems\cite{Pearson2019analog}.

Thus, both coefficient tuning and hardware noise can create situations in which the optimal solutions do not correspond to the ground states of the problem Hamiltonian but to one of its energetically proximate excited states. In quantum annealing, which is an algorithm designed to search for ground states of problem Hamiltonians, it is fundamentally incapable of finding the optimal solutions under such conditions. It should be noted that the scope of this study is not to propose a method for optimizing the coefficient tuning process itself. Instead, we focus on acquiring quantum states during the annealing process as a mitigation strategy to the aforementioned difficulties. During annealing, due to quantum fluctuation effects, the system exists in a superposition of eigenstates of the problem Hamiltonian. These quantum states at intermediate times during the annealing process, which we refer to as mid-anneal states, contain information about both ground and excited states of the problem Hamiltonian through quantum superposition.
Since the system undergoes non-adiabatic transitions during the annealing process, the population of low-lying excited states may be enhanced at intermediate times, making the acquisition of these mid-anneal states, an act we refer to in this study as mid-anneal measurement, a promising strategy for capturing optimal solutions.

Ideally, "mid-anneal measurement" refers to a non-destructive measurement that tracks the real-time evolution of the coherent quantum state without disturbing the annealing process. Such measurements are often called quantum non-demolition (QND) measurements, and theoretical protocols for them have been proposed~\cite{Sveistrys2025speeding}. Moreover, traditional destructive measurement also falls within the scope of mid-anneal measurement because it is sufficient to detect the mid-anneal state, even if the state is destroyed. Theoretical concepts for such destructive measurements are also discussed in the prior study~\cite{MunozBauza2019double-slit}. However, current hardware lacks the capability for the QND measurement, and the direct destructive measurement protocols are likewise not implemented.
Therefore, the acquisition of mid-anneal states in quantum annealing has been realized through quench. Quench, which rapidly changes the Hamiltonian during quantum annealing, has been the subject of various studies~\cite{King2018observation, Callison2021energetic, Irsigler2022quantum, King2025beyond, King2025quantum}. In practice, quench is implemented by rapidly reducing the transverse field, effectively freezing the quantum dynamics and allowing access to the instantaneous quantum state. In one of these studies, quench has been used to obtain mid-anneal states by rapidly eliminating quantum fluctuations during quantum annealing~\cite{King2018observation}. 
They employed a reverse annealing protocol in which they initialized the system in a classical state, evolved it at a specific annealing parameter, and then executed a quench by rapidly eliminating the quantum fluctuation. This operation freezes the system’s quantum state and projects the coherent state at that moment onto the computational basis. By repeatedly acquiring and sampling these snapshots at various annealing parameters (transverse field strengths), they succeeded in experimentally mapping the equilibrium properties of the Kosterlitz-Thouless (KT) phase transition.
Although such quenches are typically implemented as rapid annealing processes with finite duration, rather than truly instantaneous transitions, they still provide useful access to intermediate quantum states. However, interest in information from the quantum state during the annealing process is abundant, including in contexts of state freezing at the end of quantum annealing~\cite{Amin2015searching} and research utilizing quantum annealers as analog simulators of physical systems~\cite{King2018observation}. Therefore, with the continued development of quantum annealers, direct realization of mid-anneal state acquisition, which has been indirectly achieved through quench, can be expected. Thus, this study aims to provide foundational insights to facilitate the effective use of such future mid-anneal measurement capabilities. While our analysis focuses on theoretical foundations applicable to current quench-based approaches, the insights gained will be directly relevant to future hardware implementations with native mid-anneal measurement capabilities. Specifically, we quantitatively evaluate the effects of mid-anneal measurement on the probability of obtaining optimal and feasible solutions, and elucidate the underlying physical mechanisms that contribute to its effectiveness. Note that in constrained optimization, optimal solutions must also be feasible. Therefore, improving the probability of obtaining feasible solutions can contribute to the success rate for optimal solutions.

The remainder of this paper is organized as follows. Section~\ref{sec: theory} reviews the theoretical background of quantum annealing and the formulations of the constrained combinatorial optimization problems addressed in this study, specifically the graph bipartitioning problem and the quadratic knapsack problem. Section~\ref{Sec: method} describes our methodology, including the quantitative metric introduced to evaluate the effectiveness of mid-anneal measurement and the numerical simulation methods. Section~\ref{sec: result} presents the results of our analysis, focusing on the relationships between mid-anneal measurement effectiveness and energy structure, state similarity, and system size scaling. Section~\ref{Sec: conclusion} discusses the broader implications of our findings and their relevance to practical quantum annealing applications, and concludes the paper with a summary of our main contributions and directions for future research.

\section{Preliminary}
\label{sec: theory}

This section provides the theoretical background of this study. First, Section~\ref{Subsec: quantum annealing} outlines the basic principles of quantum annealing, a metaheuristic for solving constrained combinatorial optimization problems. Next, Section~\ref{Subsec: COP} introduces the definitions and formulations of the two constrained combinatorial optimization problems addressed in this study to evaluate the effectiveness of mid-anneal measurement: the graph bipartitioning problem (GBP) and the quadratic knapsack problem (QKP). 
These problems are chosen to represent different constraint types, such as equality constraints in GBP and inequality constraints in QKP. They also differ in the structure and dimension of their feasible solution spaces, enabling comprehensive analysis of mid-anneal measurement effectiveness across diverse problem structures.

\subsection{Quantum annealing}
\label{Subsec: quantum annealing}

Quantum annealing is a metaheuristic for solving combinatorial optimization problems. The problem Hamiltonian $H_\mathrm{c}$ is defined as:
\begin{equation}
    \label{Eq: H_c}
    H_\mathrm{c} = - \sum_{1 \leq i<j \leq N} J_{i, j} \sigma_i^z \sigma_j^z - \sum_{i=1}^N h_i \sigma_i^z.
\end{equation}
Here, $\sigma_i^z$ is the Pauli-$z$ matrix acting on spin $i$; $h_i$ is the local magnetic field acting on spin $i$; $J_{i,j}$ is the interaction between spins $i$ and $j$, and $N$ is the number of spins. We then introduce the quantum driver Hamiltonian $H_\mathrm{q}$:
\begin{equation}
    \label{Eq: H_q}
    H_\mathrm{q} = - \sum_{i=1}^N \sigma_i^x, 
\end{equation}
whose ground state is an equal superposition of all computational basis states, where $\sigma_i^x$ is the Pauli-$x$ matrix acting on spin $i$. We then construct the total time-dependent Hamiltonian $H(t)$ as follows:
\begin{equation}
    \label{Eq: total Hamiltonian of QA}
    H (t) = A(t) H_\mathrm{q} + B(t) H_\mathrm{c}, \;\;\; t \in [0, \tau],
\end{equation}
where $\tau$ is the total annealing time; $t$ is the time during the annealing process. In this annealing schedule, the system transitions from a regime dominated by quantum fluctuations, induced by the driver Hamiltonian, to one where the classical problem Hamiltonian becomes dominant. 
Here, $A(t)$ and $B(t)$ are time-dependent functions satisfying $A(0) \gg B(0)$, $A(\tau) \ll B(\tau)$, respectively. For this time-dependent Hamiltonian $H(t)$, the quantum state $\ket{\Psi(t)}$ evolves according to the Schr\"odinger equation:
\begin{equation}
    \label{Eq: Sch eq}
    \mathrm{i} \frac{\mathrm{d}}{\mathrm{d} t} \ket{\Psi(t)} = H(t) \ket{\Psi(t)}.
\end{equation}
Throughout this paper, we use natural units in which the reduced Planck constant $\hbar = 1$. When the quantum state $\ket{\Psi(t)}$ evolves adiabatically (i.e., when $\tau$ is much larger than the inverse of the minimum instantaneous energy gap), it is known to follow the instantaneous ground state of the total Hamiltonian $H(t)$~\cite{Kato1950adiabatic}. Therefore, by performing quantum annealing adiabatically by initializing the system in the ground state of the quantum driver Hamiltonian $H_\mathrm{q}$, the ground state of the problem Hamiltonian $H_\mathrm{c}$ can be obtained at $t = \tau$.

\subsection{Constrained combinatorial optimization problems}
\label{Subsec: COP}

In this study, we consider the graph bipartitioning problem (GBP) and the quadratic knapsack problem (QKP) as representative constrained combinatorial optimization problems. GBP is a problem with an equality constraint, where the number of feasible solutions can be directly controlled by the value of the constant in the constraint. Moreover, since no auxiliary variables are required, the structure of the solution space is preserved as long as the problem size is fixed. On the other hand, QKP is a problem with an inequality constraint, which necessitates the introduction of auxiliary variables to convert the constraint to equality constraints~\cite{Lucas2014ising, tanahashi2019application}, and accordingly the structure of the solution space also changes. By examining these two problems, which differ in the nature of their constraints, we aim to more comprehensively analyze how constraint characteristics influence the effectiveness of mid-anneal measurement.

\subsubsection{Graph bipartitioning problem (GBP)}
\label{sec: graph bipartitioning problem}

The graph bipartitioning problem (GBP) is to find a solution that minimizes the sum of edge weights on the partition boundary when dividing an undirected graph $G=(V,E)$ into two disjoint subgraphs $G_+=(V_+,E_+)$, $G_-=(V_-,E_-)$. Here, $V$ and $E$ are the vertex set and the edge set, respectively. As a constraint, we impose the difference between the numbers of vertices contained in each subgraph, expressed as a constant $c = \abs{V_+} - \abs{V_-}$. The value $c$ determines the size imbalance between the two partitions and is encoded by the constraint $\sum_i s_i = c$, where $s_i = \pm 1$ indicates the partition assignment. The objective function and the constraint of this problem are formulated as follows~\cite{Lucas2014ising}:
\begin{align}
    \label{Eq: Hobj of GBP classical}
    \text{minimize}& \:\:\: \frac{1}{N} \sum_{(i,j) \in E} w_{i,j} \frac{1 - s_i s_j}{2}  \;\;\left( s_i, s_j \in \{1, -1\} \right),\\
    \label{Eq: constraint of GBP classical}
    \text{subject to}& \:\:\: \sum_{i=1}^N  s_{i} = c.
\end{align}
Here, $N = \abs{V}$ is the number of vertices, and $w_{i,j}$ is the weight of edge $(i, j)$. Note that the objective function is divided by $N$ to obtain an intensive energy scale, which is appropriate since in this study we deal with fully connected problems where all edge weights $w_{i,j}$ are non-zero. Therefore, the problem Hamiltonian for the GBP is expressed as follows, based on the augmented Lagrangian function:
\begin{equation}
    \label{Eq: Hc of GBP classical}
    H_\mathrm{c} =\frac{1}{N} \sum_{(i,j) \in E} w_{i,j} \frac{1 - s_i s_j}{2} + \frac{\mu}{2} \left(\sum_{i = 1}^{N} s_i - c \right)^2
    - \lambda \left(\sum_{i = 1}^{N} s_i - c \right).
\end{equation}

\subsubsection{Quadratic knapsack problem (QKP)}
\label{subsubsec: QKP}

The quadratic knapsack problem (QKP) is to pack items into a knapsack without exceeding its weight limit, while maximizing the total sum of the values of the items in the knapsack and the compatibility between items. The objective function and the constraint of this problem are expressed as follows:
\begin{align}
    \label{Eq: Hobj of QKP classical}
    \text{minimize}& \:\:\: - \frac{1}{N} \sum_{1 \leq i \leq j \leq N} p_{i, j} x_i x_j \;\;\left( x_i, x_j \in \{0, 1\} \right),\\
    \label{Eq: constraint of QKP classical}
    \text{subject to}& \:\:\: \sum_{i=1}^N  w_{i} x_i \leq W.
\end{align}
Here, $N$ is the number of items, $p_{i,j}$ is the component of the interaction matrix, where for $i \ne j$ it represents the compatibility between items $i$ and $j$, and for $i=j$ it represents the value of item $i$. Note that we rewrite the maximization problem as a minimization of the negative objective function, which is suitable for quantum annealing frameworks. Additionally, $w_i$ is the weight of item $i$, and $W$ is the weight limit of the knapsack. To convert the inequality constraint expressed in Eq.~\eqref{Eq: constraint of QKP classical} to an equality constraint, we introduce auxiliary variables $y_d \in \{0, 1\}$. In this study, we use log encoding to reduce the number of auxiliary variables introduced. This encoding represents the total slack between the weight limit and the item weights using $D$-bit binary variables, effectively transforming the inequality into an equality constraint. This reduction in the number of variables leads to fewer required qubits and facilitates the embedding of the problem onto quantum annealing hardware. The inequality constraint expressed by Eq.~\eqref{Eq: constraint of QKP classical} is represented as follows by log encoding\cite{Jimbo2022hybrid}:
\begin{align}
    \label{Eq: constraint of QKP log}
    c(\bm{x}, \bm{y}) &\equiv \sum_{d=1}^D 2^{d-1} y_d - (2^D - 1 - W) - \sum_{i=1}^{N} w_i x_i.
\end{align}
Note that $D=\lceil \log_2W \rceil$. The constraint function $c(\bm{x}, \bm{y})$ defined in Eq.~\eqref{Eq: constraint of QKP log} is constructed such that it equals zero precisely when the original inequality constraint $\sum_{i=1}^N w_i x_i \leq W$ in Eq.~\eqref{Eq: constraint of QKP classical} is satisfied, and takes positive values otherwise. Based on this, we formulate the problem Hamiltonian $H_\mathrm{c}$ as follows, using the augmented Lagrangian function:
\begin{equation}
    H_\mathrm{c} 
    = - \frac{1}{N} \sum_{1 \leq i \leq j \leq N} p_{i, j} x_i x_j
    + \frac{\mu}{2} \left[ c (\bm{x}, \bm{y}) \right]^2 - \lambda \left[ c (\bm{x}, \bm{y}) \right].
\label{eq: Hc of QKP classical}
\end{equation}
While the QKP formulation above uses binary variables $x_i, y_d \in \{0,1\}$, it can be equivalently expressed using spin variables $s_i \in \{-1,+1\}$ through the transformation $x_i = (1+s_i)/2$, making the problem Hamiltonian in Eq.~\eqref{eq: Hc of QKP classical} directly applicable to quantum annealing.
\section{Method}
\label{Sec: method}
\begin{figure*}[t]
    \centering
    \includegraphics[width=0.8\linewidth]{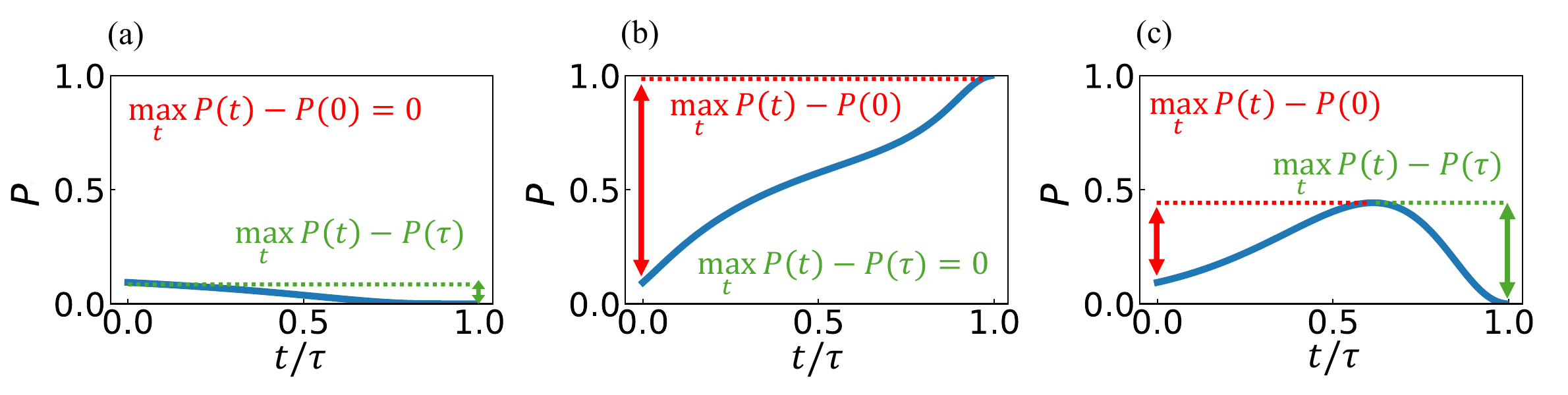}
    \caption{(Color online) Characteristics of $Q_\mathrm{d}$, which represents the effectiveness of mid-anneal measurement defined in Eq.~\eqref{Eq: Q_d}. The second factor in Eq.~\eqref{Eq: Q_d}, $\underset{t}{\mathrm{max}} P (t) - P (t=0)$, represents the improvement from the initial equal-weight superposition state. The third factor in Eq.~\eqref{Eq: Q_d}, $\underset{t}{\mathrm{max}} P (t) - P (t=\tau)$, represents the additional improvement due to mid-anneal measurement compared to standard annealing. (a), (b) Cases where mid-anneal measurement is not effective. In case (a), the second factor $\underset{t}{\mathrm{max}} P(t) - P (t=0)$ is zero, while in case (b), the third factor $\underset{t}{\mathrm{max}} P(t) - P (t=\tau)$ is zero. Therefore, $Q_\mathrm{d}$ becomes zero, indicating that mid-anneal measurement is not effective. (c) Case where mid-anneal measurement is effective. Both the first and third factors take non-zero values, resulting in a non-zero value of $Q_\mathrm{d}$, which indicates that mid-anneal measurement is effective.}
    \label{fig: Qd explanation}
\end{figure*}
This section describes the numerical simulation methods used to investigate the effectiveness of mid-anneal measurement for the problems defined in the previous section, as well as the methods for evaluating the results. First, Section~\ref{Subsec: problem setting} explains the specific parameter settings for the graph bipartitioning problem (GBP) and quadratic knapsack problem (QKP) that were the targets of our simulations, along with the computational approaches used to simulate quantum dynamics in both finite-time and the adiabatic-limit scenarios. Section~\ref{Subsec: evaluation} introduces our effectiveness metric $Q_\mathrm{d}$, establishes its theoretical foundation, and defines the energy difference quantities central to our analysis.

\subsection{Problem and simulation setting}
\label{Subsec: problem setting}

We simulated quantum annealing on the GBP and QKP instances, and on Ising models to evaluate generality. First, in Section \ref{Subsec: DeltaE-dependence of Qd}, for GBP, we set $N=6$, $0 \leq c \leq 4$, and to avoid degenerate low-energy states and ensure a dense energy spectrum suitable for evaluating mid-anneal effects, $w_{i,j}$ were generated from a uniform distribution between $0.8$ and $1.2$ for each edge $(i, j)$, meaning the graph is fully connected.
For QKP, we set $N=5$, $1 \leq W \leq 4$ with $w_{i}=1$ for all $i$, and $p_{i,j}$ were generated from a uniform distribution between $0.8$ and $1.2$ for each pair $(i, j)$.
We generated $10$ instances for each problem and performed simulations of quantum annealing under both finite time and adiabatic limit conditions. The annealing schedule used is shown below:
\begin{align}
    \label{Eq: annealing schedule of this study}
    H (t) &= A(t) H_\mathrm{q} + B(t) H_\mathrm{c},\\
    A(t) &= 1 - \frac{t}{\tau}, \\
    B(t) &= \frac{t}{\tau}.
\end{align}
For finite-time quantum annealing calculations, we computed quantum dynamics following the time-dependent Schrödinger equation in Eq.~\eqref{Eq: Sch eq} using QuTiP~\cite{Johansson2012qutip, Johansson2013qutip}, an open-source Python library. We calculated cases with annealing times $\tau = 100, 1000, 2000$. For adiabatic limit quantum annealing calculations, we obtained the ground state at each time by exact diagonalization of the instantaneous Hamiltonian $H(t)$.
Based on the time evolution of the quantum state, we calculated the probability of obtaining feasible solutions $P_\mathrm{f}(t)$, and optimal solutions $P_\mathrm{opt}(t)$ as follows:
\begin{align}
    P_\mathrm{f}(t) &= \sum_{a=1}^{N_\mathrm{f}}
    \left| \braket{\phi_{\mathrm{f}, a}}{ \Psi (t)} \right| ^2, \\
    P_\mathrm{opt}(t) &= \sum_{b=1}^{N_\mathrm{opt}}
    \left| \braket{\phi_{\mathrm{opt}, b}}{ \Psi (t)} \right| ^2.
\end{align}
Here, $N_\mathrm{f}$ and $N_\mathrm{opt}$ are the numbers of feasible and optimal solutions, respectively, and $\ket{\phi_{\mathrm{f}, a}}$ and $\ket{\phi_{\mathrm{opt}, b}}$ are their respective state vectors in the $\sigma^z$ basis.

Furthermore, to ensure consistent analysis in this study, it is important to carefully set the constraint coefficients $\lambda$ and $\mu$, which determine the structure of the problem Hamiltonian $H_\mathrm{c}$. When the constraint differs, the boundary that determines whether the ground state of $H_\mathrm{c}$ corresponds to the optimal solution changes in the $\lambda - \mu$ parameter space.
For a given $\lambda$, as $\mu$ increases from small values, there exists a boundary where the ground states of the problem Hamiltonian begin to correspond to optimal solutions. We define $\mu^*$ as the value of $\mu$ at this boundary, which represents the minimum penalty coefficient required to ensure that optimal solutions become the ground states of $H_\mathrm{c}$. Note that $\mu^*$ is a function of $\lambda$, as different values of $\lambda$ shift this boundary.
To ensure consistency across all analyses in this study, we set $\lambda$ such that $\mu^* = 1$ for all problem instances. This normalization ensures that all problem instances are analyzed at the same relative distance from the feasibility boundary, enabling fair comparison of mid-anneal measurement effectiveness across different constraint configurations.

In Section \ref{Subsec: Discussion}, we considered the Ising model to investigate the generality of the physical mechanisms that make mid-anneal measurement effective. The problem Hamiltonian is given by the following Ising model:
\begin{equation}
    \label{Eq: ising model}
    H_\mathrm{c} = -  \sum_{1 \leq i<j \leq N} \frac{J_{i, j}}{N} \sigma_i^z \sigma_j^z - \sum_{i=1}^N h_i \sigma_i^z.
\end{equation}
The factor $1/N$ is introduced because in fully connected systems there are $\mathcal{O}(N^2)$ interaction terms, and this normalization ensures that the energy scale remains finite as the system size becomes large. Here, we consider two types of models: ferromagnetic (FM) and antiferromagnetic (AF) models. For each pair of spins $(i,j)$, the interactions $J_{i,j}$ were randomly set from a uniform distribution $J_{i,j} \in [0, 1]$ for the ferromagnetic model and from $J_{i,j} \in [-1, 0]$ for the antiferromagnetic model. Furthermore, to reduce the effects of system symmetry, the local magnetic fields $h_i$ were randomly generated from a uniform distribution $h_i \in [0, 2]$ for both models.
In Section \ref{subsubsec: ising N dependence}, we applied efficient numerical calculation methods utilizing the symmetry of the system for fully connected Ising models with uniform interactions and performed analysis on large-scale systems. Such large-scale system analysis is important for evaluating the practical applicability of the proposed method in quantum annealing. 
For uniform systems, the Ising model in Eq.~\eqref{Eq: ising model} can be simplified as follows:
\begin{equation}
    \label{eq: ising model uniform}
    H_\mathrm{c} = - \frac{J}{N} \sum_{1 \leq i<j \leq N} \sigma_i^z \sigma_j^z - h \sum_{i=1}^N \sigma_i^z.
\end{equation}
Here, $J$ is the uniform interaction and $h$ is the uniform local magnetic field. The characteristic of this system is that all spins are equivalent, making it possible to represent the state space in terms of total spin bases rather than individual spin configurations. Specifically, we can utilize the total spin operators that describe the total magnetization of the entire system. The total $z$-component magnetization operator $S^z$ represents the sum of all individual $z$-components, while the total $x$-component magnetization operator $S^x$ represents the sum of all individual $x$-components. These operators act on the collective Hibert space and have eigenvalues ranging from $-N/2$ to $+N/2$ in steps of $1$, resulting in $(N+1)$ distinct eigenvalues.
By expressing the Hamiltonian in terms of these total spin operators, the problem Hamiltonian $H_\mathrm{c}$ in Eq.~\eqref{eq: ising model uniform} can be rewritten as follows:
\begin{equation}
    \label{eq: ising model with total spin operator}
    H_\mathrm{c} = - \frac{J(S^z)^2}{2N} - hS^z + \text{const}.
\end{equation}
This representation allows us to reduce the dimension of the matrix representing the Hamiltonian from $2^N$ to $(N+1)$, enabling efficient computation for large-scale problems.

\subsection{Evaluation}
\label{Subsec: evaluation}

We introduce $Q_\mathrm{d}$ as a metric to quantitatively evaluate the effectiveness of mid-anneal measurement. $Q_\mathrm{d}$ is calculated from the time evolution of probability. Let $P(t)$ denote the probability of obtaining desired solutions such as feasible solutions. $Q_\mathrm{d}$ is defined as follows using the initial value $P(t=0)$, maximum value $\underset{t}{\mathrm{max}} P (t)$, and final value $P(t=\tau)$:
\begin{align}
    \label{Eq: Q_d}
    Q_\mathrm{d} &\equiv \chi \left[ \max_t P (t) - P(t=0)\right] \left[ \max_t P (t) - P(t=\tau)\right], \\
    \label{eq: chi}
    \chi &= \frac{1}{1 - P(t = 0)}.
\end{align}
$Q_\mathrm{d}$ is normalized such that its value lies in $[0,1]$.
Here, $\chi$ is a normalization constant introduced to suppress the dependence on the initial probability $P(t = 0)$.

Figure \ref{fig: Qd explanation} explains the validity of $Q_\mathrm{d}$ as a metric representing the effectiveness of mid-anneal measurement. In this study, we assume that by performing mid-anneal measurement during solution search, we can obtain a state that achieves the maximum probability $\underset{t}{\mathrm{max}} P (t)$.
The second factor in Eq.~\eqref{Eq: Q_d}, $\underset{t}{\mathrm{max}} P (t) - P (t=0)$, represents the improvement in probability due to annealing from the initial time $t=0$ (when all states are obtained with equal probability since quantum fluctuation is dominant). That is, if $\underset{t}{\mathrm{max}} P (t) - P(t=0) = 0$, it means that annealing is not effective. This factor allows us to quantitatively measure the effect of annealing.
On the other hand, the third factor, $\underset{t}{\mathrm{max}} P (t) - P(t=\tau)$, represents the improvement in probability compared to a standard annealing schedule that performs solution search until the end $t = \tau$. If $\underset{t}{\mathrm{max}} P (t) = P(t=\tau)$, it implies that standard annealing can obtain solutions as good as or better than mid-anneal measurement. Therefore, we consider cases where both $\underset{t}{\mathrm{max}} P (t) - P (t=0)$ and $\underset{t}{\mathrm{max}} P (t) - P(t=\tau)$ have high values as cases where mid-anneal measurement is effective.
Note that $Q_\mathrm{d}$ takes values in the range $[0, 1]$, where $Q_\mathrm{d} = 0$ indicates no advantage from mid-anneal measurement, and $Q_\mathrm{d} = 1$ represents the ideal case where both improvements are maximized.
Based on this general definition of $Q_\mathrm{d}$, we calculate specific metrics from the time evolution of the probability of obtaining feasible solutions $P_\mathrm{f} (t)$ and optimal solutions $P_\mathrm{opt} (t)$, denoted as $Q_\mathrm{d,f}$ and $Q_\mathrm{d,opt}$, respectively.

Furthermore, to understand the relationship between the solution space structure in quantum annealing and the effect of mid-anneal measurement, we define the energy difference between feasible and infeasible solutions $\Delta E_\mathrm{f}$ as follows:
\begin{equation}
    \label{Eq: Delta_E_f}
    \Delta E_\mathrm{f} \equiv E_\mathrm{infeasible, min} - E_\mathrm{feasible, min}.
\end{equation}
Here, $E_\mathrm{infeasible, min}$ is the minimum energy of infeasible solutions and $E_\mathrm{feasible, min}$ is that of feasible solutions. This energy difference $\Delta E_\mathrm{f}$ is defined with respect to the energy structure of the problem Hamiltonian $H_\mathrm{c}$.
The value of $\Delta E_\mathrm{f}$ is determined by the coefficients of constraint terms $\mu, \lambda$, and is an important quantity that characterizes the relationship between the energy levels of feasible and infeasible solutions in the problem Hamiltonian. Therefore, in this study, we primarily investigate the dependence on $\Delta E_\mathrm{f}$ to analyze the relationship between the effectiveness of mid-anneal measurement and the structure of the solution space.

\section{Results}
\label{sec: result}
\begin{figure}[t]
    \begin{minipage}[h]{1\linewidth}
        \centering
        \includegraphics[scale=0.5]{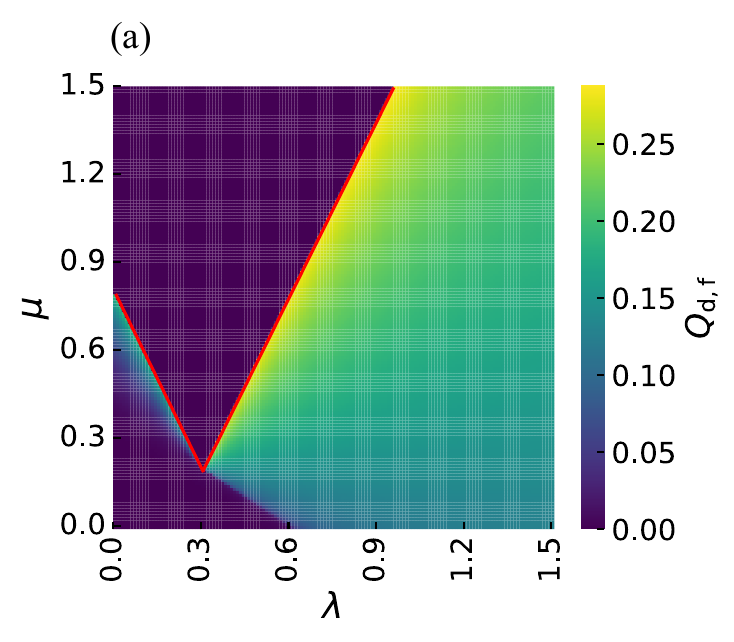}
    \end{minipage}
    \begin{minipage}[h]{1\linewidth}
        \centering
        \includegraphics[scale=0.5]{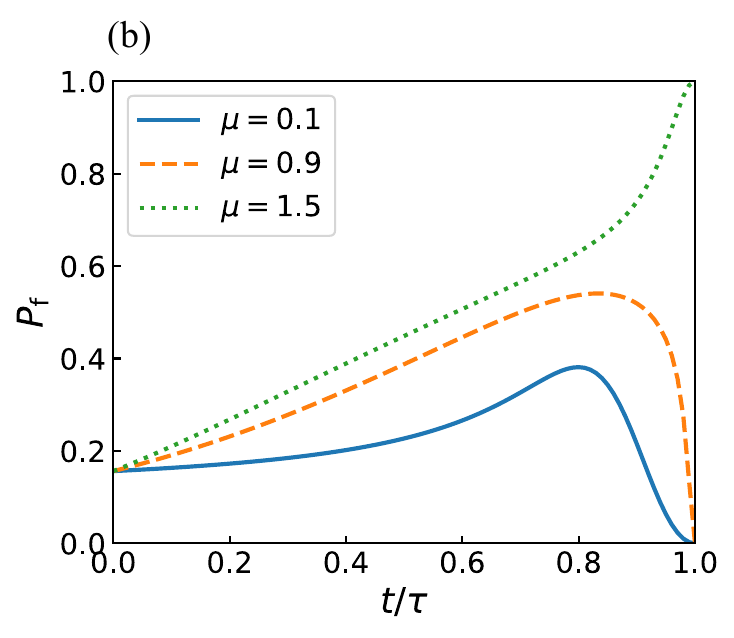}
    \end{minipage}
\caption{(Color online) Example of the relationship between the effectiveness of mid-anneal measurement and constraint coefficients. These results are from quantum annealing simulations in the adiabatic limit for QKP, $N=5$, $W=1$. (a) Dependence of $Q_\mathrm{d, f}$ on $\lambda$ and $\mu$. The red line indicates the theoretical boundary in the $\lambda - \mu$ space where the ground state of $H_\mathrm{c}$ switches from feasible to infeasible. (b) Adiabatic time evolution of the probability of obtaining feasible solutions $P_\mathrm{f}(t)$ for different values of $\mu$ at $\lambda=0.7$ (where $\mu^* = 1$).}
\label{fig: mu-lam map}
\end{figure}
\begin{figure}[t]
    \centering
    \includegraphics[scale=0.5]{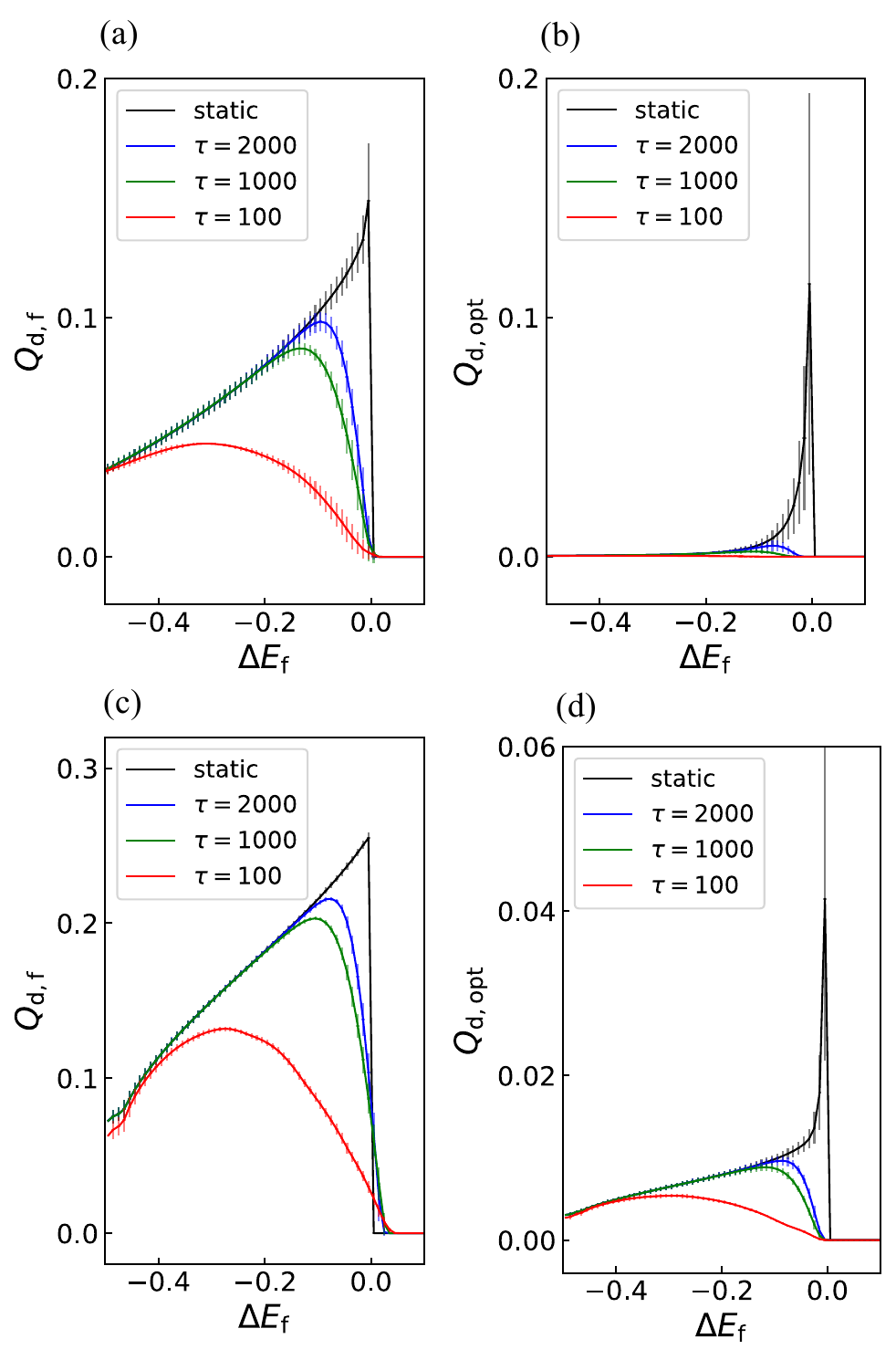}
    \caption{(Color online) Dependence of the effectiveness of mid-anneal measurement $Q_\mathrm{d, f}$, $Q_\mathrm{d, opt}$ on the energy difference between the minimum energies of feasible and infeasible solutions $\Delta E_\mathrm{f}$. Parameters are: $N=6$, $c=0$, $\lambda=-0.916$ for GBP; and  $N=5$, $W=1$, $\lambda=0.7$ for QKP. Results are shown for quantum annealing in the adiabatic limit (static) and with annealing time $\tau$. Calculations were performed for 10 instances, with $\Delta E_\mathrm{f}$ grouped in intervals of $0.01$. The mean (solid lines) and standard deviation (error bars) were then calculated. (a) GBP, $Q_\mathrm{d, f}$; (b) GBP, $Q_\mathrm{d, opt}$; (c) QKP, $Q_\mathrm{d, f}$; (d) QKP, $Q_\mathrm{d, opt}$.}
\label{fig: Delta_Ef dependence}
\end{figure}
\begin{figure}[t]
    \centering
    \includegraphics[scale=0.5]{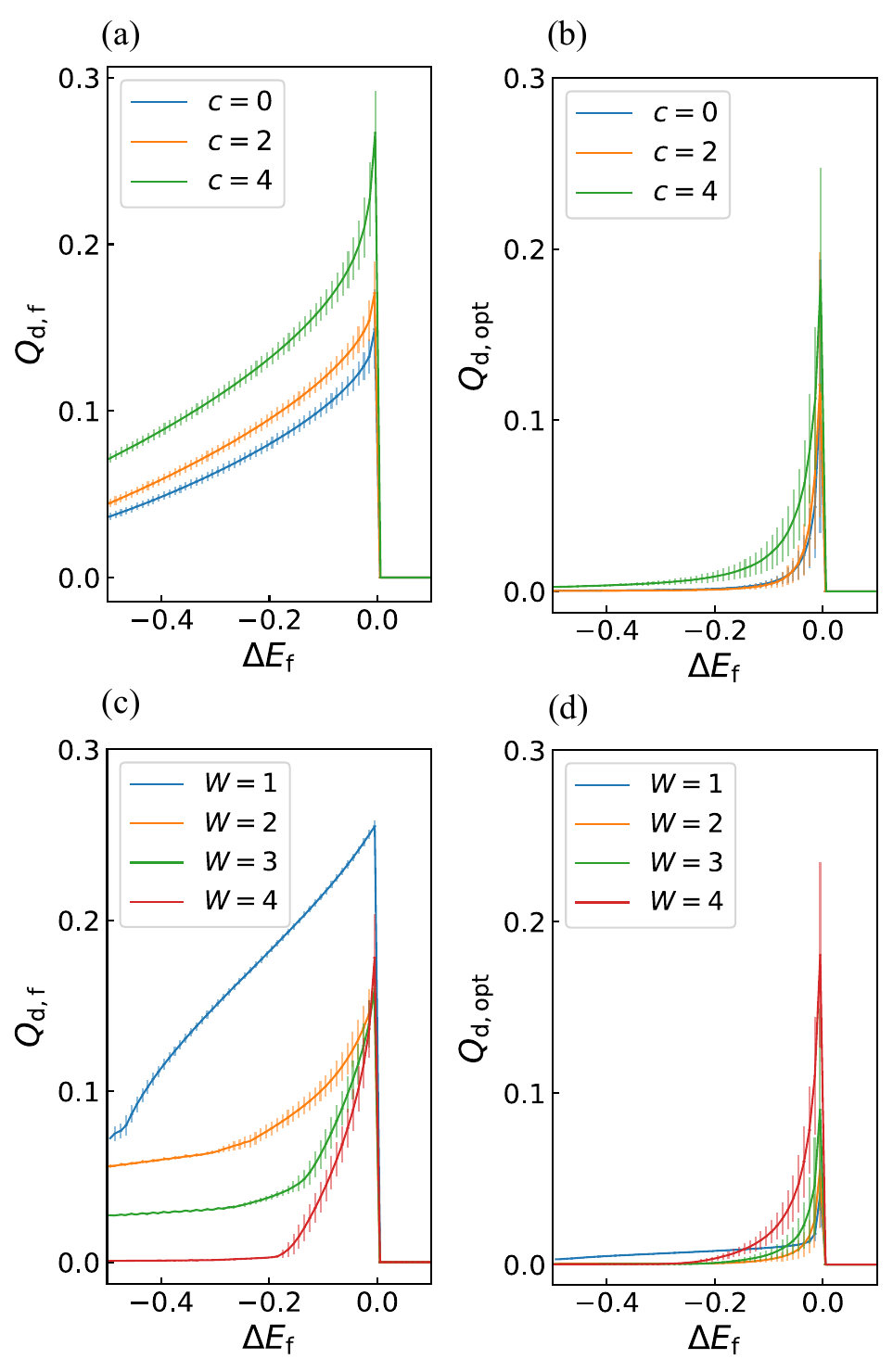}
\caption{(Color online) Dependence of the effectiveness of mid-anneal measurement $Q_\mathrm{d, f}$, $Q_\mathrm{d, opt}$ on the energy difference between the minimum energies of feasible and infeasible solutions $\Delta E_\mathrm{f}$. Parameters are: $N=6$ for GBP, $N=5$ for QKP. Results are from quantum annealing simulations in the adiabatic limit, comparing the effects of partitioning constraint $c$ in GBP and weight constraint $W$ in QKP. Calculations were performed for 10 instances, with $\Delta E_\mathrm{f}$ grouped in intervals of 0.01. The mean (solid lines) and standard deviation (error bars) were then calculated. $\lambda$ was set such that $\mu^*=1$. (a) GBP, $Q_\mathrm{d, f}$; (b) GBP, $Q_\mathrm{d, opt}$; (c) QKP, $Q_\mathrm{d, f}$; (d) QKP, $Q_\mathrm{d, opt}$.}
\label{fig: constraint dependence}
\end{figure}
\begin{figure}[t]
    \centering
    \includegraphics[scale=0.5]{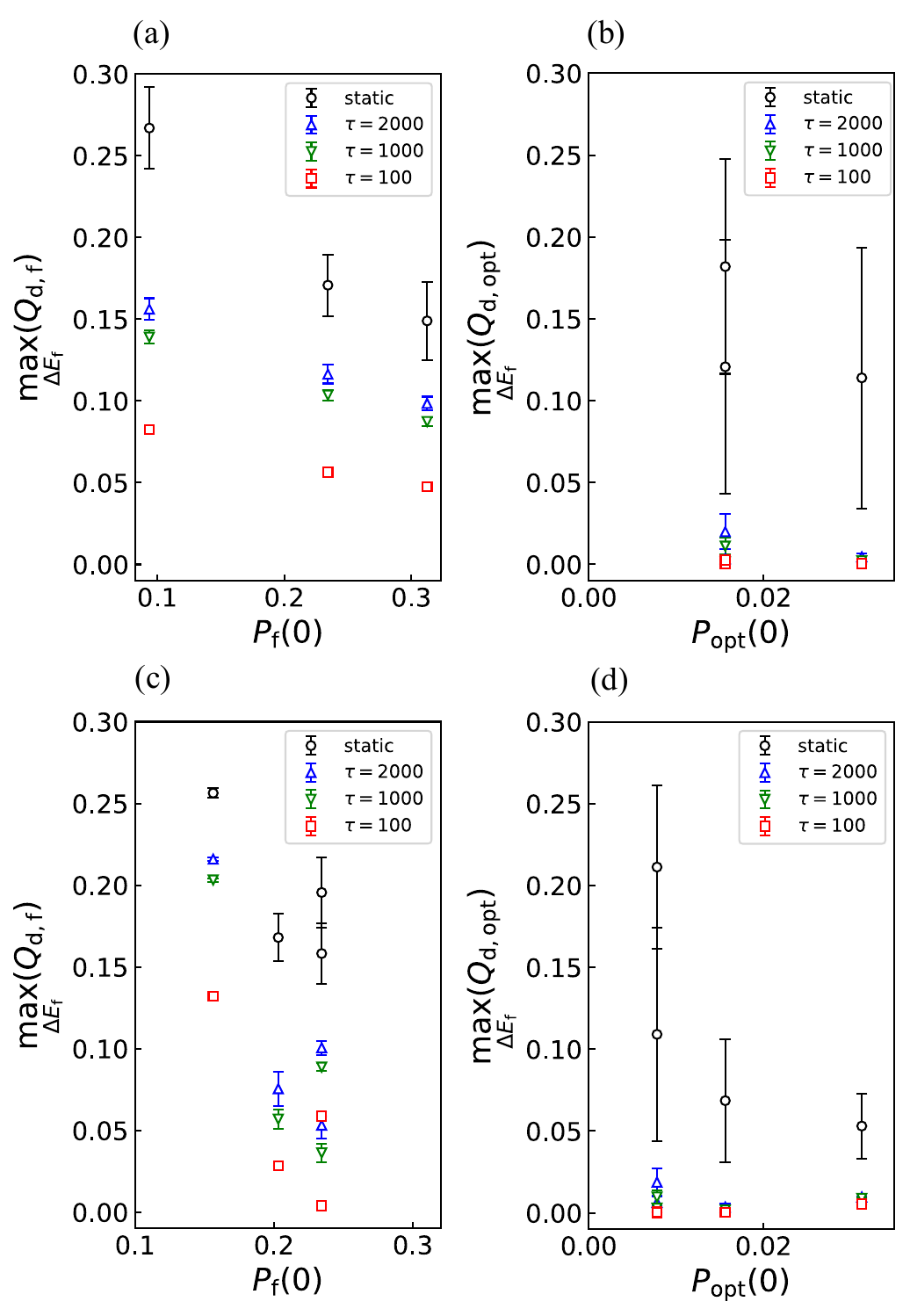}
    \caption{(Color online) Relationship between the effectiveness of mid-anneal measurement and the probability at the initial time of annealing. The vertical axis plots the maximum values of $Q_\mathrm{d, f}$ and $Q_\mathrm{d, opt}$ when $\Delta E_\mathrm{f}$ is varied for each instance, denoted as $\underset{\Delta E_\mathrm{f}}{\mathrm{max}} Q_\mathrm{d, f}$, $\underset{\Delta E_\mathrm{f}}{\mathrm{max}} Q_\mathrm{d, opt}$, which represent the potential for mid-anneal measurement to function most effectively. The results are from quantum annealing simulations in the adiabatic limit. The mean (solid lines) and standard deviation (error bars) were calculated over $10$ instances. The horizontal axis shows the initial probabilities $P_\mathrm{f}(0)$ for $Q_\mathrm{d, f}$ and $P_\mathrm{opt}(0)$ for $Q_\mathrm{d, opt}$. (a) GBP, $P_\mathrm{f}(0)$ vs $\underset{\Delta E_\mathrm{f}}{\mathrm{max}} Q_\mathrm{d, f}$; (b) GBP, $P_\mathrm{opt}(0)$ vs $\underset{\Delta E_\mathrm{f}}{\mathrm{max}} Q_\mathrm{d, opt}$; (c) QKP, $P_\mathrm{f}(0)$ vs $\underset{\Delta E_\mathrm{f}}{\mathrm{max}} Q_\mathrm{d, f}$; (d) QKP, $P_\mathrm{opt}(0)$ vs $\underset{\Delta E_\mathrm{f}}{\mathrm{max}} Q_\mathrm{d, opt}$.}
\label{fig: P(0) vs Q_d}
\end{figure}
%

This section presents simulation results evaluating the effectiveness of mid-anneal measurement, using the metrics $Q_\mathrm{d,f}$ and $Q_\mathrm{d,opt}$ defined in Section~\ref{Subsec: evaluation}. This section is organized as follows. First, Section~\ref{Subsec: DeltaE-dependence of Qd} reports the results of investigating the effectiveness of mid-anneal measurement for GBP and QKP, both of which are constrained combinatorial optimization problems. In this subsection, we particularly clarify the relationship between the metrics $Q_\mathrm{d, f}$, $Q_\mathrm{d, opt}$ and the energy difference between feasible and infeasible solutions, $\Delta E_\mathrm{f}$. The subsequent Section~\ref{Subsec: Discussion} discusses the generality of the physical mechanisms that contribute to the effectiveness of mid-anneal measurement, building upon the findings obtained in Section~\ref{Subsec: DeltaE-dependence of Qd}. For this purpose, we analyze the effectiveness of mid-anneal measurement for basic Ising models and consider the physical background that makes mid-anneal measurement effective in quantum annealing.

\subsection{Energy difference and constraint dependence of $Q_\mathrm{d, f}$, $Q_\mathrm{d, opt}$}
\label{Subsec: DeltaE-dependence of Qd}

First, we begin by overviewing the dependence of the constraint coefficients $\lambda$ and $\mu$ on the effectiveness of mid-anneal measurement. For this purpose, we take up QKP and show the results of adiabatic quantum annealing, evaluating $Q_\mathrm{d, f}$ in the $\lambda-\mu$ space, as presented in Fig.~\ref{fig: mu-lam map}.
In Fig.~\ref{fig: mu-lam map}(a), regions exist in the $\lambda-\mu$ space where the effectiveness of mid-anneal measurement $Q_\mathrm{d, f}$ exhibits locally large values. These regions correspond to areas at which $\mu$ is slightly smaller than the boundary (red line in Fig.~\ref{fig: mu-lam map}(a)) where the ground states of the problem Hamiltonian $H_\mathrm{c}$ switch from feasible to infeasible solutions. In such parameter regions, the feasible solutions are encoded not in the ground states of $H_\mathrm{c}$ but in low-energy excited states, and the energy difference between these excited states and the ground states is very small. In this situation, as shown in the case of $\mu=0.9$ ($\mu < \mu^*$) in Fig.~\ref{fig: mu-lam map}(b), the peak value of the probability of obtaining the feasible solutions $P_\mathrm{f}(t)$ becomes large. Therefore, by performing mid-anneal measurement at the peak time, the feasible solutions can be obtained with higher probability than when continuing annealing to the end, resulting in a large value of $Q_\mathrm{d, f}$.
On the other hand, when deviating from these regions, $Q_\mathrm{d, f}$ takes on small values. In regions where $\mu$ is larger than the boundary ($\mu > \mu^*$), feasible solutions are correctly encoded in the ground states of $H_\mathrm{c}$, so the standard annealing method is effective, and $Q_\mathrm{d,f}$ becomes zero. Conversely, in regions where $\mu$ is very small, the penalty imposed on infeasible solutions is too small. As a result, the peak value of $P_\mathrm{f}(t)$ becomes lower, and $Q_\mathrm{d,f}$ remains small. Note that in Fig.~\ref{fig: mu-lam map}(a), there exist regions where $Q_\mathrm{d, f}$ remains small despite $\mu$ being smaller than the boundary. This corresponds to situations where numerous infeasible solutions are encoded near the ground state, causing the impact of feasible solutions to diminish.

Next, to further generalize the behavior of mid-anneal measurement effectiveness in this $\lambda-\mu$ space and provide a physical quantitative evaluation, we analyze the dependence of $Q_\mathrm{d, f}$ and $Q_\mathrm{d, opt}$ on the energy difference $\Delta E_\mathrm{f}$. Figure \ref{fig: Delta_Ef dependence} shows the result of simulations with different annealing times $\tau$ for two problems: GBP ($N=6$, $c=0$) and QKP ($N=5$, $W=1$).
From Fig.~\ref{fig: Delta_Ef dependence}, it became clear that the effect of mid-anneal measurement strongly depends on the properties of the ground state of the problem Hamiltonian $H_\mathrm{c}$. 
When $\Delta E_\mathrm{f} < 0$, that is, when the ground state of $H_\mathrm{c}$ becomes an infeasible solution, the effectiveness of mid-anneal measurement $Q_\mathrm{d, f}$, $Q_\mathrm{d, opt}$ is found to increase. 
On the other hand, in the region where $\Delta E_\mathrm{f} > 0$, the effect of mid-anneal measurement becomes negligible or vanishes in most cases. This phenomenon can be theoretically understood from the fact that in adiabatic quantum annealing, the ground state of $H_\mathrm{c}$ is inevitably obtained at time $\tau$. Under such circumstances, it is not effective to conduct a measurement during the annealing process to obtain mid-anneal states. 
A particularly noteworthy point is that in adiabatic quantum annealing, in the region where $\Delta E_\mathrm{f} < 0$, $Q_\mathrm{d, f}$ and $Q_\mathrm{d, opt}$ monotonically increase as $\abs{\Delta E_\mathrm{f}}$ becomes small. This behavior can be understood from the fact that when $\abs{\Delta E_\mathrm{f}}$ is small, the impact of excited states of $H_\mathrm{c}$ appears strongly in the annealing process. Since this tendency was confirmed from the results of the adiabatic limit simulations, it represents an essential property of quantum annealing. 
Moreover, the results for finite annealing times are generally based on this fundamental property observed in the adiabatic limit. However, finite-time dynamics introduce oscillatory behavior in the time evolution of $P_\mathrm{f} (t)$ and $P_\mathrm{opt} (t)$, which can lead to small non-zero values of $Q_\mathrm{d,f}$ and $Q_\mathrm{d,opt}$ even in regions where $\Delta E_\mathrm{f} > 0$ in some instances, such as Fig.~\ref{fig: Delta_Ef dependence}(c).
As described above, even if the ground state of $H_\mathrm{c}$ is not the desired solution (a feasible/optimal solution), if the energy level of that solution is close to the ground state, the probability of acquiring the desired solution can be improved through mid-anneal measurement. This trend was also observed when formulated based on the penalty function as shown in Fig.~\ref{fig: Delta_Ef dependence penalty} in the appendix.

This finding has important implications for practical application of quantum annealing to constrained combinatorial optimization problems. In particular, mid-anneal measurement enables the acquisition of desired solutions even when they are encoded in excited states of the problem Hamiltonian. This suggests the possibility of facilitating coefficient tuning in cases where appropriate adjustment of constraint coefficients $\mu$ and $\lambda$ is difficult or when hardware noise becomes significant.

Next, to quantitatively analyze the relationship between constraints and the effectiveness of mid-anneal measurement, we investigated the dependence of $Q_\mathrm{d, f}$ and $Q_\mathrm{d, opt}$ on the constraints $c$ and $W$. Figure~\ref{fig: constraint dependence} shows the $\Delta E_\mathrm{f}$ dependence of $Q_\mathrm{d, f}$ and $Q_\mathrm{d, opt}$ under different constraint settings. Here, to understand the essential properties of the system, we focused on quantum annealing in the adiabatic limit.
From Fig.~\ref{fig: constraint dependence}, it can be observed that in GBP, $Q_\mathrm{d, f}$ and $Q_\mathrm{d, opt}$ tend to show larger values for larger values of the constant $c$ representing the partitioning constraint of the undirected graph. On the other hand, in QKP, $Q_\mathrm{d, f}$ shows a negative correlation with the weight constraint $W$, while $Q_\mathrm{d, opt}$ shows a positive correlation.

To understand the origin of these characteristics, we focused on the probabilities of obtaining feasible solutions and optimal solutions in the initial annealing state, $P_\mathrm{f}(0)$ and $P_\mathrm{opt}(0)$. Here, $P_\mathrm{f}(0)$ and $P_\mathrm{opt}(0)$ are important quantities that characterize the structure of the solution space. These represent the probabilities of obtaining feasible solutions and optimal solutions in the initial state where all states are superposed with equal weights. In other words, these values directly reflect the existence ratios of feasible solutions and optimal solutions in the entire solution space. Therefore, $P_\mathrm{f}(0)$ and $P_\mathrm{opt}(0)$ serve as proxies for the relative size of the feasible and optimal subspaces within the full Hilbert space.

Figure \ref{fig: P(0) vs Q_d} shows the relationship between the maximum effectiveness of mid-anneal measurement $\underset{\Delta E_\mathrm{f}}{\mathrm{max}} Q_\mathrm{d}$ and the initial probabilities. Note that the initial probabilities, $P_\mathrm{f}(0)$ and $P_\mathrm{opt}(0)$, can take the same values even for different constraint settings, as these probabilities simply represent the ratio of the number of feasible or optimal solutions to the total number of possible solutions.
From Fig.~\ref{fig: P(0) vs Q_d}, a slight negative correlation was observed between $Q_\mathrm{d, f}$ and $P_\mathrm{f}(0)$, as well as between $Q_\mathrm{d, opt}$ and $P_\mathrm{opt}(0)$. As a possible interpretation of these trends, this suggests that mid-anneal measurement becomes effective when the proportion of desired solutions (feasible or optimal solutions) in the entire solution space is small. 
However, it should be noted that $P_\mathrm{f}(0)$ and $P_\mathrm{opt}(0)$ approach zero in the limit of $N \rightarrow \infty$. Therefore, for very large problem sizes, this dependence on $P_\mathrm{f}(0)$ and $P_\mathrm{opt}(0)$, and consequently the dependence on constraints $c$ and $W$ observed in Fig.~\ref{fig: constraint dependence} may not appear. This implies that the constraint dependence observed in our results might be attributed to the relatively small system sizes studied.

\subsection{Discussion}
\label{Subsec: Discussion}

In this section, we provide a deeper analysis of the results obtained for constrained combinatorial optimization problems in the previous section and examine the generality of the physical mechanisms that make mid-anneal measurement effective. For this purpose, we investigated the effectiveness of mid-anneal measurement using the Ising model expressed by Eq.~\eqref{Eq: ising model}, which serves as a simpler and more universal model.
In Section~\ref{Subsec: DeltaE-dependence of Qd}, we confirmed that mid-anneal measurement becomes effective when the desired solutions (optimal or feasible solutions) correspond to excited states rather than the ground states of the problem Hamiltonian $H_\mathrm{c}$. Based on this finding, in this discussion, we regarded the first excited state of the Ising model as the desired solution, and calculated the effectiveness of mid-anneal measurement $Q_\mathrm{d, e1}$ according to Eq.~\eqref{Eq: Q_d} from the time evolution of its acquisition probability, $P_\mathrm{e1}(t)$. The simulations were performed as quantum annealing in the adiabatic limit using the annealing schedule defined in Eq.~\eqref{Eq: annealing schedule of this study}.

\subsubsection{Energy difference dependence}
\label{subsubsec: ising energy difference dependence}

\begin{figure}[t]
    \begin{minipage}[h]{1\linewidth}
        \centering
        \includegraphics[scale=0.5]{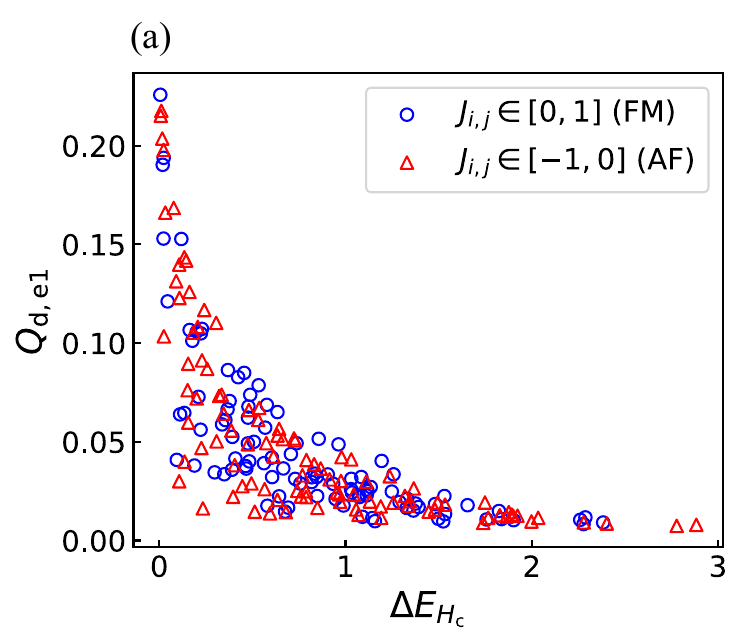}
    \end{minipage}
    \begin{minipage}[h]{1\linewidth}
        \centering
        \includegraphics[scale=0.5]{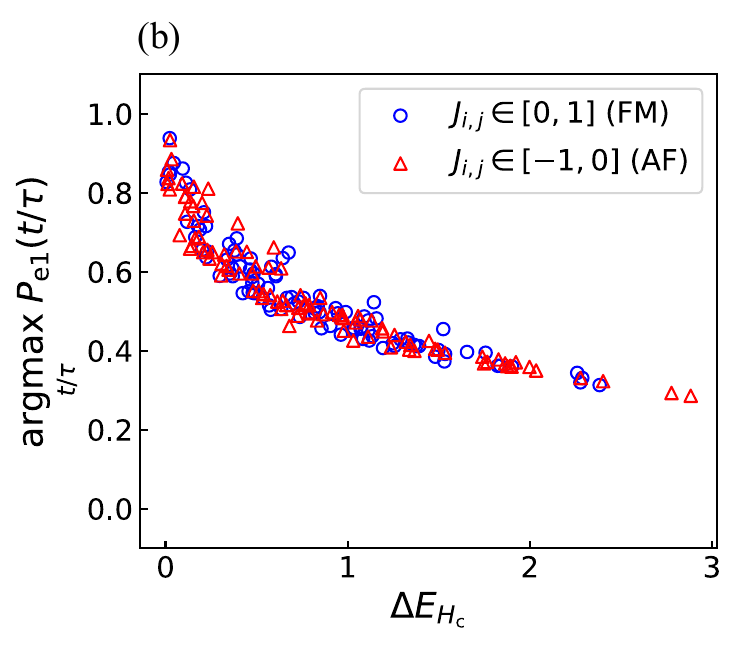}
    \end{minipage}
\caption{(Color online) Dependence on $\Delta E_{H_\mathrm{c}}$. Results are from quantum annealing simulations in the adiabatic limit for $N=4$ and $h_i \in [0,2]$ over $100$ instances. 
(a) Effectiveness of mid-anneal measurement $Q_\mathrm{d, e1}$; (b) Optimal mid-anneal measurement timing, $\underset{t/ \tau}{\mathrm{argmax}} P_\mathrm{e1} (t / \tau)$.}
\label{fig: Delta E_Hc dependence}
\end{figure}

In this section, we present the dependence on the energy difference of the problem Hamiltonian of Ising model expressed by Eq.~\eqref{Eq: ising model}. In Section~\ref{Subsec: DeltaE-dependence of Qd}, a strong dependence was observed on the energy difference $\Delta E_\mathrm{f}$ between the minimum energy of feasible solutions and that of infeasible solutions. To examine the correspondence with this result, here we analyze the dependence on the energy difference $\Delta E_{H_\mathrm{c}}$ between the ground state and the first excited state of the Ising model. $\Delta E_{H_\mathrm{c}}$ is defined as follows:
\begin{equation}
    \label{Eq: delta_E_Hc}
    \Delta E_{H_\mathrm{c}} = E_\mathrm{e1} - E_\mathrm{gs}.
\end{equation}
Here, $E_\mathrm{gs}$ and $E_\mathrm{e1}$ are the energies of the ground state and the first excited state of the Ising model, respectively. Figure~\ref{fig: Delta E_Hc dependence}(a) shows the $\Delta E_{H_\mathrm{c}}$ dependence of $Q_\mathrm{d, e1}$. We also investigated the $\Delta E_{H_\mathrm{c}}$ dependence of the optimal mid-anneal measurement timing during the quantum annealing process (i.e., the timing when $P_\mathrm{e1}(t)$ takes its maximum value), $\underset{t/ \tau}{\mathrm{argmax}} P_\mathrm{e1} (t / \tau)$.
From Fig.~\ref{fig: Delta E_Hc dependence}(a), it can be seen that mid-anneal measurement becomes effective in more cases when $\Delta E_{H_\mathrm{c}}$ is smaller. This trend is consistent with the results confirmed in Section~\ref{Subsec: DeltaE-dependence of Qd}, which showed a negative correlation with $|\Delta E_\mathrm{f}|$. This suggests that the energy difference of the problem Hamiltonian is closely related to the effectiveness of mid-anneal measurement. Furthermore, similar dependence on $\Delta E_{H_\mathrm{c}}$ is observed in both ferromagnetic (FM) and antiferromagnetic (AF) models. From Fig.~\ref{fig: Delta E_Hc dependence}(b), it can be seen that the optimal mid-anneal measurement timing becomes earlier as $\Delta E_{H_\mathrm{c}}$ becomes large, appearing to follow a hyperbolic relationship. However, while this specific functional form collapses when parameters are varied, a consistent negative correlation is robustly observed across different parameter settings. This can be attributed to the fact that when $\Delta E_{H_\mathrm{c}}$ is large relative to quantum fluctuations, the contribution from excited states diminishes at an early stage of quantum annealing, making the ground state of the problem Hamiltonian dominant.

\subsubsection{$N$ dependence}
\label{subsubsec: ising N dependence}

\begin{figure}[t]
    \centering
    \includegraphics[scale=0.5]{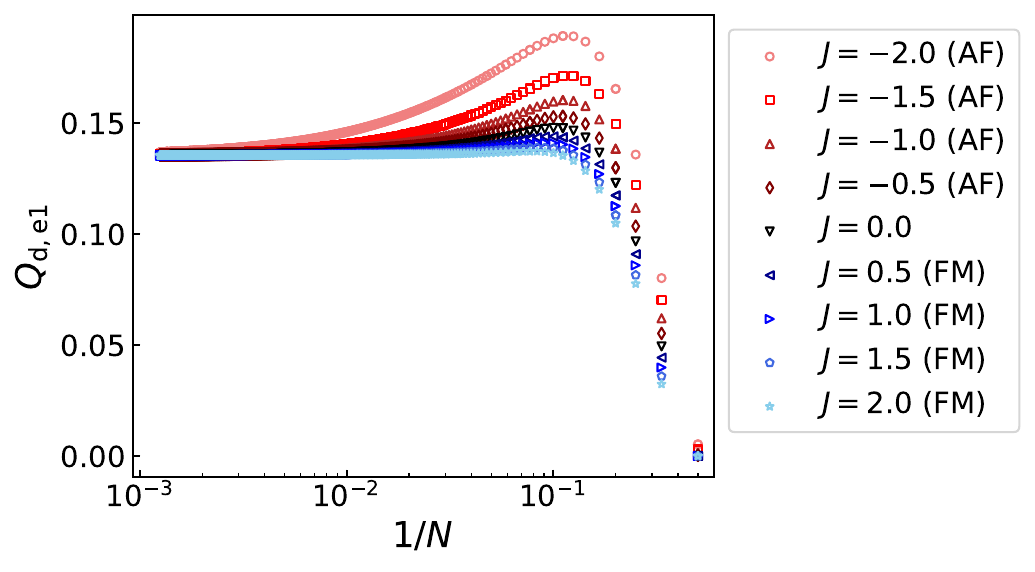}
\caption{(Color online) Dependence of the effectiveness of mid-anneal measurement, $Q_\mathrm{d, e1}$, on $1/N$. Results are from quantum annealing simulations in the adiabatic limit for $h=2.5$.}
\label{fig: N dependence}
\end{figure}

Having generalized our findings from GBP and QKP in the previous section, in this section, we investigate a crucial question for practical applicability: how the effectiveness of mid-anneal measurement scales with the system size $N$. To efficiently examine this scalability in large-scale systems, we target the fully connected Ising model with uniform interactions described by Eq.~\eqref{eq: ising model uniform}. As noted in Section~\ref{Subsec: problem setting}, this model's symmetry permits efficient computation for large $N$.
The local magnetic field was fixed at $h = 2.5$ to eliminate trivial degeneracy and to ensure that the density of low-energy states remains comparable across all system sizes. Figure~\ref{fig: N dependence} shows the $1/N$ dependence of $Q_\mathrm{d, e1}$ in this model.
From Fig.~\ref{fig: N dependence}, it can be seen that the effectiveness of mid-anneal measurement, $Q_\mathrm{d, e1}$, takes non-zero values even for large $N$. Additionally, an interesting point is that a tendency for $Q_\mathrm{d, e1}$ to asymptotically approach a constant value with increasing $N$ is observed. The cause of this behavior is a subject for future research.

\section{Conclusion}
\label{Sec: conclusion}

This study proposed mid-anneal measurement as a mitigation strategy for addressing the difficulties of coefficient tuning and hardware noise in quantum annealing applied to constrained combinatorial optimization problems. Specifically, these factors often lead to situations where the true optimal solution fails to be encoded as the ground state of problem Hamiltonian. We posited that mid-anneal measurement can effectively resolve this critical issue by enabling access to such solutions residing in excited states. 
We also aimed to quantitatively evaluate its effectiveness while elucidating the physical mechanisms that make mid-anneal measurement effective. To this end, we first introduced a metric, $Q_\mathrm{d}$, to quantify the effectiveness of mid-anneal measurement. Using this metric, we conducted numerical simulations on the graph bipartitioning problem (GBP) and the quadratic knapsack problem (QKP), both representative constrained combinatorial optimization problems, as well as on Ising models, and obtained the following conclusions.

The main achievements of this study are twofold.
First, we identified energy proximity as the key factor for mid-anneal measurement effectiveness. Specifically, we confirmed that mid-anneal measurement is most effective when the desired solutions (feasible or optimal) correspond not to the ground states of $H_\mathrm{c}$, but to low-energy excited states that are energetically close to them.

Second, we established the scalability of the mid-anneal measurement method. Through simulations on fully connected Ising models, we observed that the effectiveness of mid-anneal measurement persists even as the system size increases, with $Q_\mathrm{d}$ values converging to a finite constant. Furthermore, enhanced effects in antiferromagnetic systems suggest that this approach is especially useful in harder instances where ground-state search is inherently difficult.

These findings provide important implications for the practical application of quantum annealing. Mid-anneal measurement can serve as a practical mitigation approach for obtaining better solutions in situations where the optimal solutions are not encoded as the ground states of the problem Hamiltonian due to difficulties in adjusting constraint coefficients or hardware noise. When actually using quantum annealers, it is impossible to determine externally whether a failure in encoding the optimal solutions as the ground states has occurred. Therefore, an appropriate strategy is to alternate between standard quantum annealing and mid-anneal measurement across multiple sampling runs. This hybrid approach can mitigate the risk of failing to encode the optimal solution in the ground state.

Future work includes further theoretical investigation into how problem complexity and structure influence mid-anneal measurement effectiveness. In particular, investigating the relationship between problem difficulty (such as frustration in spin systems) and mid-anneal measurement effectiveness could provide deeper insights into the applicability of this approach. 
It should be noted that our analysis focuses on the principal effectiveness of mid-anneal measurement against static noise (such as $J$-chaos) that alters the problem Hamiltonian's energy structure. Our simulations are based on ideal quantum dynamics and do not consider dynamic noise sources, such as quantum state decoherence or thermal fluctuations. Investigating how these dynamic noise processes impact the effectiveness is a crucial study.
Additionally, experimental validation on real quantum annealers will be critical. Also, the mid-anneal measurement approach may serve as a tool for generating diverse solutions, such as Pareto optimal solutions in multi-objective optimization problems.

\section*{acknowledgement}
This work was partially supported by the Japan Society for the Promotion of Science (JSPS) KAKENHI (Grant Number JP23H05447), the Council for Science, Technology, and Innovation (CSTI) through the Cross-ministerial Strategic Innovation Promotion Program (SIP), ``Promoting the application of advanced quantum technology platforms to social issues'' (Funding agency: QST), Japan Science and Technology Agency (JST) (Grant Number JPMJPF2221). The computations in this work were partially performed using the facilities of the Supercomputer Center, the Institute for Solid State Physics, The University of Tokyo. S. Tanaka wishes to express their gratitude to the World Premier International Research Center Initiative (WPI), MEXT, Japan, for their support of the Human Biology-Microbiome-Quantum Research Center (Bio2Q).

\bibliographystyle{jpsj}
\bibliography{reference}
\appendix
\section{Results on penalty function}
\label{sec: appendix}

\begin{figure}[h]
    \centering
    \includegraphics[scale=0.5]{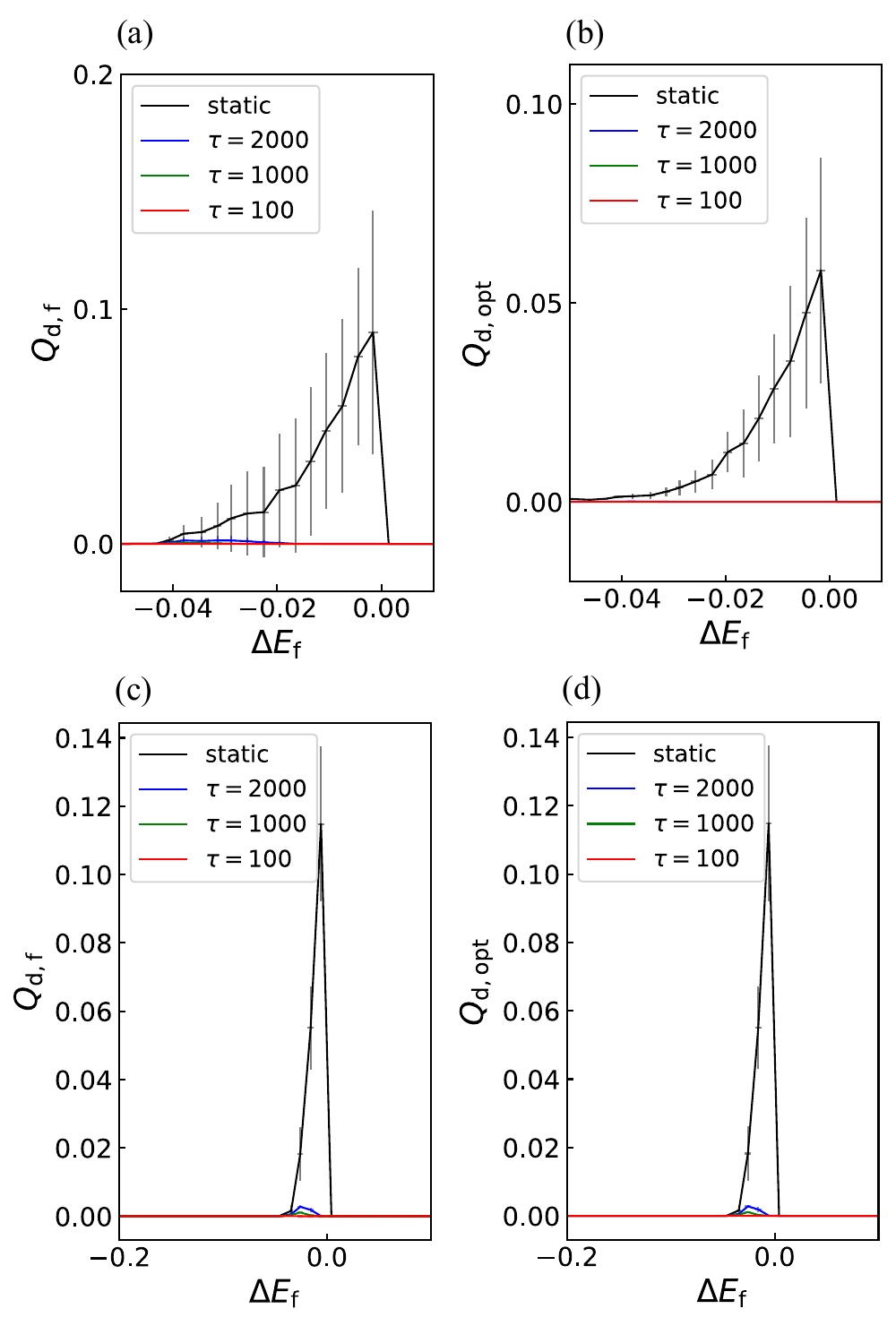}
    \caption{(Color online) Dependence of the effectiveness of mid-anneal measurement $Q_\mathrm{d, f}$, $Q_\mathrm{d, opt}$ on the energy difference between the minimum energies of feasible and infeasible solutions $\Delta E_\mathrm{f}$. Formulations of the problems are based on penalty function. Parameters are: $N=6$, $c=0$ for GBP; and  $N=5$, $W=1$ for QKP. Results are shown for quantum annealing in the adiabatic limit (static) and with annealing time $\tau$. Calculations were performed for 10 instances, with $\Delta E_\mathrm{f}$ grouped in intervals of $0.002$. The mean (solid lines) and standard deviation (error bars) were then calculated. (a) GBP, $Q_\mathrm{d, f}$; (b) GBP, $Q_\mathrm{d, opt}$; (c) QKP, $Q_\mathrm{d, f}$; (d) QKP, $Q_\mathrm{d, opt}$.}
\label{fig: Delta_Ef dependence penalty}
\end{figure}

In this study, we conducted investigations based on augmented Lagrangian functions. On the other hand, penalty functions are also widely used for formulating constrained combinatorial optimization problems. Therefore, we investigated whether the trends confirmed in this study are also observed for penalty functions. 
To this end, we analyzed the dependence of $Q_\mathrm{d, f}$ and $Q_\mathrm{d, opt}$ on the energy difference $\Delta E_\mathrm{f}$. The investigation was conducted under the same settings as in Fig.~\ref{fig: Delta_Ef dependence}, which are $N=6$, $c=0$ for GBP, $N=5$, $W=1$ for QKP.
Figure~\ref{fig: Delta_Ef dependence penalty} suggests that the dependence on $\Delta E_\mathrm{f}$ is preserved even when using penalty functions. However, the effectiveness metrics $Q_\mathrm{d, f}$ and $Q_\mathrm{d, opt}$ appear to exhibit substantially reduced values, particularly in non-adiabatic scenarios. This reduction could be attributed to the possibility that penalty functions may yield more restrictive parameter regions where mid-anneal measurement becomes effective, with these regions possibly exhibiting stronger dependence on the specific constraint. These preliminary observations suggest that mid-anneal measurement might demonstrate enhanced utility when applied to augmented Lagrangian formulations compared to penalty function approaches. However, a more comprehensive comparison between penalty and augmented Lagrangian formulations should be conducted in future work to establish definitive conclusions regarding their relative effectiveness.

\end{document}